# Estimation of missing data by using the filtering process in a time series modeling

## Ahmad Mahir R. and Al-khazaleh A. M. H.

*School of Mathematical Sciences*
*Faculty of Science and Technology*
*Universiti Kebangsaan Malaysia*
*43600 UKM Selangor D. E.,*
*MALAYSIA*
*e-mail:* `mahir@ukm.my`
*e-mail:* `ahmed_2005kh@yahoo.com`

**Abstract:** This paper proposed a new method to estimate the missing data by using the filtering process. We used datasets without missing data and randomly missing data to evaluate the new method of estimation by using the Box - Jenkins modeling technique to predict monthly average rainfall for site 5504035 Lahar Ikan Mati at Kepala Batas, P. Pinang station in Malaysia. The rainfall data was collected from the $1^{st}$ January 1969 to $31^{st}$ December 1997 in the station. The data used in the development of the model to predict rainfall were represented by an autoregressive integrated moving - average (ARIMA) model. The model for both datasets was ARIMA$(1,0,0)(0,1,1)_s$.The result checked with the Naive test, which is the Thiel's statistic and was found to be equal to $U = 0.72086$ for the complete data and $U = 0.726352$ for the missing data, which mean they were good models.

**Keywords and phrases:** ARIMA model, monthly average rainfall, filtering process and, forecasting method.

## 1. Introduction

Time series is a set of observations recorded over a time. Autoregressive integrated Moving Average models or ARIMA models are especially suited to short-term forecasting because most ARIMA models place heavy emphasis on the recent past rather than the distant past. An observed series theoretically consists of two parts: the first part is the series generated by real process, and the second noise which is the result of outside disturbances. Elimination of this noise is the main aim of a time series analysis. Early developments to eliminate noise came from introducing the autoregressive approach and moving average approach (ARIMA). The Box -Jenkins procedure consists of the implementation or completion of several steps, or stages: identification, estimation, diagnostic checking and forecasting. In the identification of an appropriate Box -Jenkins model: changing the data, if necessary, into a stationary time series and determining the tentative model by analyzing the autocorrelation and partial autocorrelation function (2; 5). The estimation for the constant and the coefficients of the equation must be obtained. The main purpose of this investigation is to analyze the data collected automatically and to evaluate a predictive model and then produce a set of forecast for site at which the data was collected. According to Pankratz, in his study, Box - Jenkins method produced the best forecast for 74% of the series that he evaluated (4) .The cost associated with the Box - Jenkins approach in a given situation is generally greater than many other quantitative methods. The Box - Jenkins model is the most general way of approaching forecasting and unlike other models, there is no need to assume, initially a fixed pattern and it is not limited to specific kind of pattern. These models can be fitted to any set of time series data by selecting the appropriate value of the parameters p, d, q to suit individual series. A problem frequently encountered in data collection is a missing observations or observations may be virtually impossible to obtain, either because of time or cost constrains. In order to replace that observations, there are several different options available to the researchers. Firstly, replace with the mean of the series. Secondly replace with the nave forecast. Also replace with a simple trend forecast. Finally replace with an average of the last two known observations that bound the missing observations.

## 2. Description of the data set

The rainfall data was collected from the $1^{st}$ January 1969 to $31^{st}$ December 1997 in the station. In this research the data on rainfall amount were collected and recorded daily. The monthly averages were calculated by finding the sum of all the amount of rainfall in that particular month and divide it by the number of days in that month for each year.

## 3. Methodology for missing observations

A problem frequently encountered in data collection is a missing observation in a data series. In order to replace that observation, there are several different options available to the researchers. Firstly, replace with the mean of the series. This mean can be calculated over the entire range of the sample. Secondly, replace with the naive forecast. Naive model is the simplest form of a Univariate forecast model, this model uses the current time period's value for the next time period, that is $\hat{Y}_{t+1} = Y_t$. Also, replace with a simple trend forecast. This is accomplished by estimating the regression equation of the form $Y_t = a + b.t$ (where t is the time) for the periods prior to the missing value. Then use the equation to fit the time periods missing. Finally, replace with an average of the last two known observations that bound the missing observations. This paper suggested new method to estimates the missing data by using the filtering process (1). The filtering process is:

$$y_t = \frac{1}{\sum w'_{i+1}} \sum_{i=0}^{M} w'_{i+1} x_{t-i} = \sum_{i=0}^{M} w_{i+1} x_{t-i} \qquad (1)$$

where $w_{i+1} = w'_{i+1} / \sum w'_{i+1}$ is the weight and M is the number of observations in a moving average. We substitute $w'_{i-1} = \varphi^i$ where $\varphi$ is the correlation of the entire data. Therefore, the corresponding moving average is

$$y_t = \varphi x_t + \varphi^2 x_{t-1} + ... + \varphi^M x_{t-m}. \qquad (2)$$

where $x_t$ is the original observations. We transformed the complete data by using equation (2) and we built an appropriate model. After that we assumed there are holes spaced randomly in the data. If $y_s$ were missed (where s is index of the hole), we substituted the average of the complete data instead of $x_s$ then we calculate the future value

$$y_s = \varphi \bar{y} + \varphi^2 x_{s-1} + ... + \varphi^M x_{s-m}. \qquad (3)$$

Then we built the model for the data that contained the holes. We applied the same model on the new data. We compared the result of the model for the two datasets by using Box-Jenkins ARIMA model in the next section.

## 4. Box - Jenkins ARIMA models

The Box - Jenkins method is a procedure for accomplishing the model past values of the time series variable and past values of the error terms. The Box - Jenkins approach consists of extracting the predictable from the observed data through a series of iterations. The most common ARIMA model included three parameters: p, d, and q where p is the number of autoregressive parameters, d is the number of differencing parameters and q is the number of moving average parameters. A general ARIMA model is in the form:

$$z_t = C + \varphi_1 z_{t-1} + \varphi_2 z_{t-2} + ... + \varphi_p z_{t-p} + a_t - \theta_1 a_{t-1} - ... - \theta_q a_{t-q}. \qquad (4)$$

where:
$t$: is the periodic time
$z_t$ : is the numerical value of an observation
$\varphi_i$: for $i = 1, 2, ...p$ are the autoregressive parameters
$\theta_j$ for $j = 1, 2, ..., q$ are the moving average parameters
$a_t$: is the shock element at time $t$

To estimate the parameters $\varphi_i$ and $\theta_j$ for a fixed p and q we perform the linear multiple regression

$$\hat{z}_t = \mu + \varphi_1 z_{t-1} + \varphi_2 z_{t-2} + ... + \varphi_p z_{t-p} - \theta_1 a_{t-1} - ... - \theta_q a_{t-q}. \qquad (5)$$

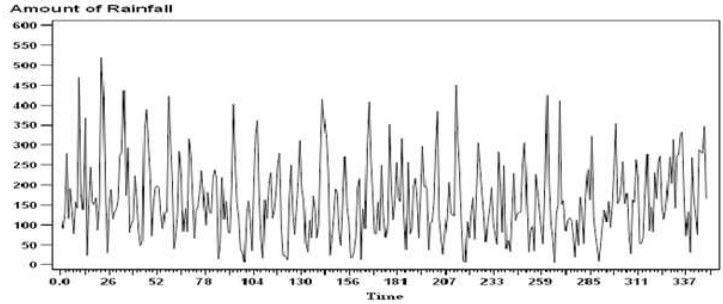

Fɪɢ 1. *Plot of the original data*

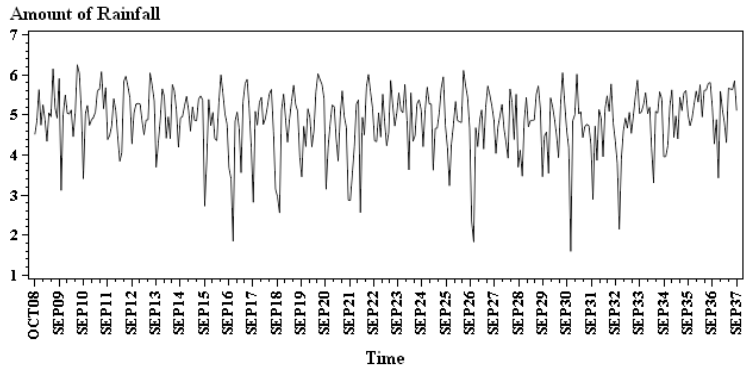

Fɪɢ 2. *Plot of the Transform data*

There are two phases to the identification of an appropriate Box - Jenkins model: changing the data if necessary into a stationary time series and determining the tentative model by observing the behavior of the autocorrelation and partial autocorrelation function. A stationary time series is that it does not contain trend, that is, it fluctuates around a constant mean. By looking at a time series plot (see figure 1 plot of data without transformation and difference). The rainfall data in Pinang was in need of a transformation. By taking logarithm it will transform the series into a stationary time series as can be seen in the figure (2). The first differencing was for seasonal part by subtraction the values of two adjacent observations in the series that is,$z_t = \Delta Y_t = Y_t - Y_{t-12}$ for seasonal. We can write the differencing by the operator of differencing as the following $B z_t = z_{t-1}$ . After transformation, it is clear that the observations fluctuate around the constant mean. Box and Jenkins suggest the number of Lag to be no more than $n/4$ autocorrelations, the autocorrelation coefficient measures the correlation between a set of observations and a lagged set of observation in a time series. The autocorrelation between $z_t$ and $z_{t+k}$ measures the correlation between the pairs $(z_1, z_{1+k}), (z_2, z_{2+k}), ...., (z_n, z_{n+k})$ The sample autocorrelation

| Parameter | Estimate | Standard Error | t Value | Approx Pr > |t| | Lag |
|-----------|----------|----------------|---------|----------------|-----|
| $MA(1,1)$ | 0.85667 | 0.02923 | 29.31 | <0.0001 | 12 |
| $AR(1,1)$ | 0.15889 | 0.05410 | 2.94 | 0.0035 | 1 |

coefficients $r_k$ is an estimate of $\rho_k$ where

$$r_k = \frac{\sum (z_t - \overline{z})(z_{t+k} - \overline{z})}{\sum (z_t - \overline{z})^2}. \qquad (6)$$

with
$z_t$:the data from the stationary time series.
$z_{t+k}$ : the data from $k$ time period ahead of $t$
$\overline{z}$: the mean of the stationary time series.
The estimated partial autocorrelation function PACF is used as a guide, along with the estimated autocorrelation function ACF, in choosing one or more ARIMA models that might fit the available data. The idea of partial autocorrelation analysis is that we want to measure how $\hat{z}_t$ and $\hat{z}_{t+k}$ are related. The equation that gives a good estimate of the partial autocorrelation is

$$\hat{\varphi}_{kj} = \hat{\varphi}_{k-1,j} - \hat{\varphi}_{kk}\hat{\varphi}_{k-1,k-j}. \qquad (7)$$

for $k = 3, 4, ...; j = 1, 2, ..., k-1$ We can find the shape of the ACF and PACF in a seasonal model as you see in the figures (3) and (4). So, the multiplicative seasonal ARIMA model $(p, d, q) \times (P, D, Q)_s$ is a generalization and is considered as an extension of the method to series in which a patterns repeat seasonally over time, where the parameters $(p, d, q)$ are for no seasonal and the parameters $(P, D, Q)_s$ are for the seasonal parts. Once a stationary time series has been selected (the ACF cuts off or dies down quickly), we can identify a tentative model by examining the behavior of the ACF and PACF. In the mixed model both the ACF and PACF dies down exponentially. The figures of the ACF and PACF as you in see in figures (3) and (4)

## 5. Results

The t-statistics as you can see in Table 1 and also in Table 2 .associated with $\Theta_{12}$ and $\varphi$ are greater in absolute value than 2, therefore, indicating that these parameters should be retained in the model for both datasets.

We deduced from the foregoing tables the first model for the complete data is

$$z_t = 0.159z_{t-1} - a_t + 0.857a_{t-12}$$

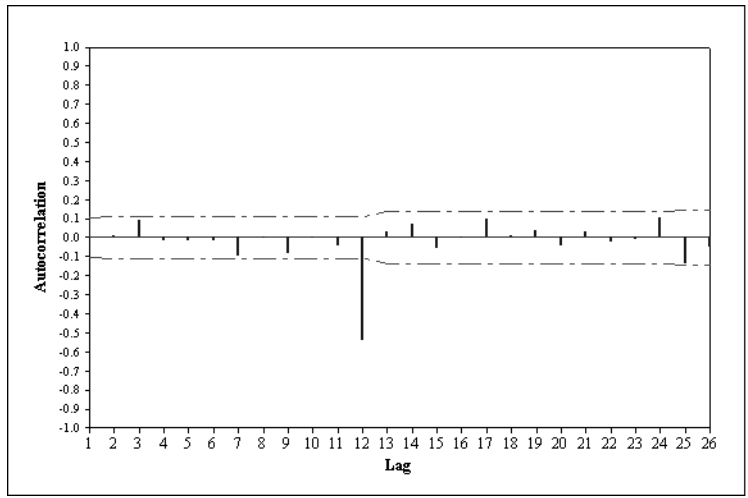

Fɪɢ 3. *Autocorrelation Function*

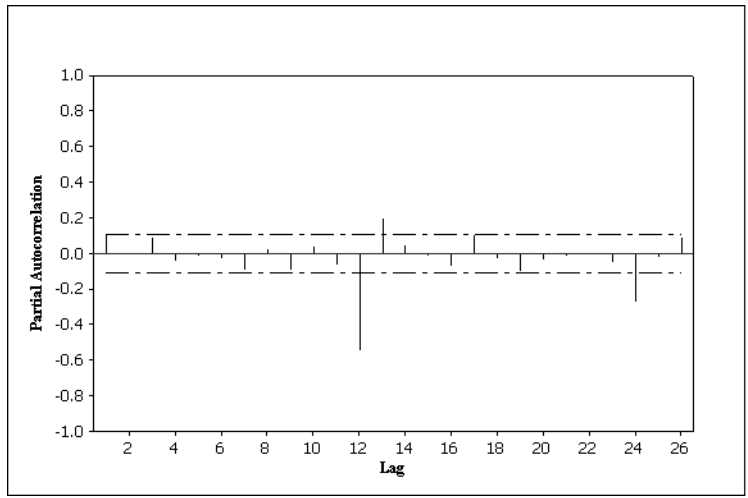

Fɪɢ 4. *Partial Autocorrelation Function*

Tᴀʙʟᴇ 2
*Parameters for missing data*

| Parameter | Estimate | Standard Error | t Value | Approx Pr > \|t\| | Lag |
|-----------|----------|----------------|---------|-------------------|-----|
| $MA(1,1)$ | 0.85383 | 0.02969 | 28.76 | <0.0001 | 12 |
| $AR(1,1)$ | 0.17818 | 0.05390 | 3.31 | 0.0010 | 1 |

and the model with missing data is $z_t = 0.178z_{t-1} - a_t + 0.854a_{t-12}$ At the estimation stage, we get the precise estimates of a small number of parameters. Then tentatively we choose an ARIMA $(1, 0, 0)(0, 1, 1)_s$ model. We fit these models to the data to get precise estimate of parameters: $\varphi_1$ for non seasonal AR part, and $\Theta_{12}$ for MA coefficient for seasonal parameter. We dropped the mean $\mu$ from the model since the mean of working series is -0.00207 and the standard deviation is 0.897664 for the complete data and for the dataset with missing data the mean is -0.00051 and the standard deviation is 0.889339. Also we note that since the first value of $z_t$ that can be calculated is $z_{13} = z_{13}^* - z_1^*$ where b=13 since the t-test $\frac{\bar{z}}{s_z/\sqrt{n-b+1}} = \frac{-0.00207}{0.897664/\sqrt{432-13+1}} = -0.04726$ which is less than 2 for the complete data. The t-test $\frac{\bar{z}}{s_z/\sqrt{n-b+1}} = \frac{-0.00051}{0.889339/\sqrt{432-13+1}} = 0.01175$ which is less than 2 for the missing data. We conclude that $\bar{z}$ is statistically close to zero and that it should be omitted from the model for the two datasets.

## 6. Diagnostic checking

At the diagnostic checking stage, we used the Ljung-Box statistic (denoted by $Q^*$ as in Equation (8) to check the adequacy of the model by examining the autocorrelation and partial autocorrelation of the residuals (2; 6).

$$Q^* = n'(n' + 2) \sum_{l=1}^{K} (n' - l)^{-1} r_l^2(a'). \qquad (8)$$

here $n' = (n - d)$ where n is the number of observations in the original time series, $r_l(a')$ is the sample autocorrelation of the residuals at lag $l$ and d is the degree of non seasonal differencing used to transform the original time series values into stationary time series values. The p-values associated with $Q^*$ indicate that the model

$$z_t = 0.159z_{t-1} - a_t + 0.857a_{t-12}$$

is adequate for the complete data since the p-value is greater than 0.05 and less than the chi square for values of K equal 6,12, 18, 24 and 36 . For example, since d=0 is the degree of differencing for the non seasonal, the $n'$ used to calculate $Q^*$ is

$$n' = (n - d) = 432 - 0 = 432$$

Therefore, if we let K=6,

$$Q^* = n'(n' + 2) \sum_{L=1}^{6} (n' - L)^{-1} r_l^2(a') = (336)(336 + 2) \left[ (336 - 1)^{-1} (-.01576)^2 \right.$$
$$+ (336 - 2)^{-1}(0.08733)^2 + (336 - 3)^{-1}(0.08257)^2 + (336 - 4)^{-1}(.04915)^2 +$$
$$(336 - 5)^{-1}(0.00493)^2 + (336 - 6)^{-1} (0.0019)^2 \left. \right] = 5.8385165$$

We use the rejection point $\chi^2_{[\alpha]}(K - 0) = \chi^2_{[0.05]}(6) = 12.5916$ since $Q^* = 5.84 < 12.5916$ , we cannot reject the adequacy of the model by setting $\alpha = 0.05$ .

Table 3
*Parameters for complete data*

| Lag | $Q^*$ | D.F | $P > Q^*$ | Autocorrelations |
|-----|-------|-----|-----------|------------------|
| 6   | 5.84  | 4   | 0.2116    | -0.016 0.087 0.083 -0.049 0.005 -0.002 |
| 12  | 11.77 | 10  | 0.3004    | -0.069 0.021 -0.023 0.054 0.065 -0.064 |
| 18  | 23.41 | 16  | 0.1032    | 0.087 0.089 0.020 -0.001 0.124 -0.039 |
| 24  | 29.00 | 22  | 0.1448    | 0.033 0.010 0.050 0.064 -0.064 0.060 |
| 30  | 36.54 | 28  | 0.1292    | -0.097 -0.011 0.093 -0.037 0.018 -0.028 |
| 36  | 40.38 | 34  | 0.2089    | 0.017 0.068 -0.006 0.064 0.027 -0.022 |

Table 4
*Parameters for complete data*

| Lag | $Q^*$ | D.F | $P > Q^*$ | Autocorrelations |
|-----|-------|-----|-----------|------------------|
| 6   | 4.70  | 4   | 0.3191    | -0.016 0.076 0.071 -0.045 0.015 -0.023 |
| 12  | 13.02 | 10  | 0.2227    | -0.076 0.004 -0.049 0.092 0.048 -0.070 |
| 18  | 26.77 | 16  | 0.0441    | 0.073 0.114 -0.009 -0.033 0.135 -0.035 |
| 24  | 31.58 | 22  | 0.0848    | 0.030 -0.004 0.022 0.067 -0.019 0.084 |
| 30  | 37.08 | 28  | 0.1172    | -0.094 -0.012 0.062 0.008 0.032 -0.033 |
| 36  | 45.83 | 34  | 0.0847    | 0.039 0.104 -0.002 0.091 -0.022 -0.047 |

The p-value is the area under the curve of the chi-square distribution having 5 degrees of freedom to the right of $Q^* = 5.84$ Also the p-value is 0.2116. Since $p - value = 0.2116 > 0.05 = \alpha$ , we cannot reject the adequacy of the model by setting $\alpha = 0.05$ This demonstrates that comparing the p-value with $\alpha$ yields the same conclusion as comparing $Q^*$ with $\chi^2_{[\alpha]}(K - n_c)$ . However tables (3) and (4) show that the p-value associated with $Q^*$ for K=6, 12, 18, 24, 30 and 36 are all greater than 0.05 for the two datasets with and without missing data, and there are no spikes in the plot of the autocorrelation of residual figures (5) and (6), we conclude that the model is adequate. Similarly we got the $Q^*$ for the model with missing data.

In order to forecast the natural logarithm of the monthly amount of rainfall in the next 2 years (months 337 through 349), we note that since $z_t = y'_t - y'_{t-12}$

where $y'_t = \ln y_t$ we can express the model for the complete data as

$$y'_t = y'_{t-12} + 0.159(y'_{t-1} - y'_{t-13}) - a_t + 0.857a_{t-12}$$

, and the model

$$y'_t = y'_{t-12} + 0.178(y'_{t-1} - y'_{t-13}) - a_t + 0.854a_{t-12}$$

for the data with holes. By using the least squares point estimates; these forecasts are shown in table (5). Several models were examined. Results of estimating monthly average of rainfall forecasting with 95% confidence interval were presented in the table (5).

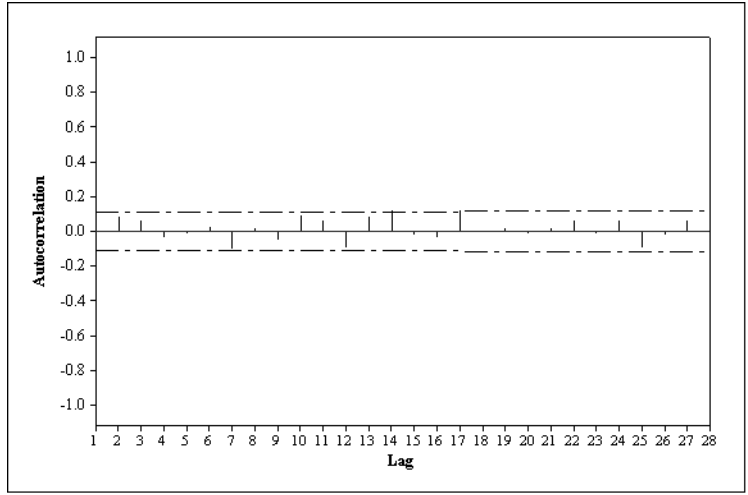

FIG 5. *Autocorrelation Plot of Residuals for the complete data*

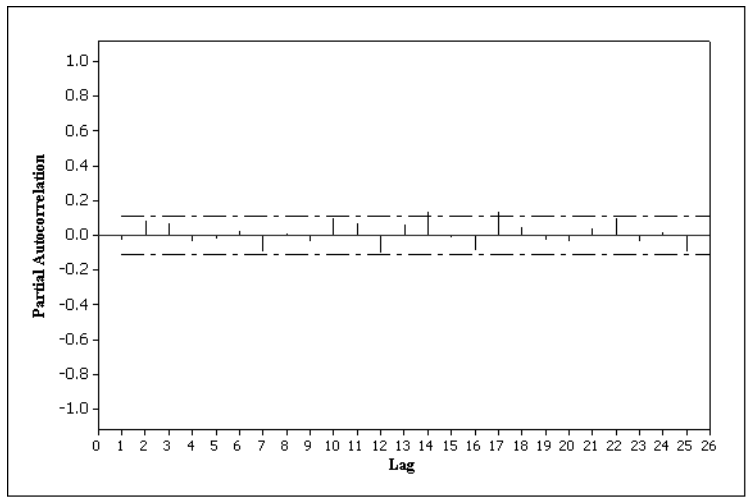

FIG 6. *Partial Autocorrelation Plot of Residuals for the complete data*

TABLE 5
*Parameters for Complete Data*

| Obs | F.C.D | 95% C. D. | | F. M.D | 95% C. D. | |
|-----|-------|-------|-------|--------|-------|-------|
| 349 | 4.1241 | 2.8177 | 5.4305 | 4.2168 | 2.9284 | 5.5052 |
| 350 | 4.1645 | 2.8417 | 5.4873 | 4.1798 | 2.8711 | 5.4884 |
| 351 | 4.5674 | 3.2442 | 5.8907 | 4.5886 | 3.2792 | 5.8979 |
| 352 | 5.2507 | 3.9275 | 6.5740 | 5.2552 | 3.9458 | 6.5645 |
| 353 | 5.0672 | 3.7439 | 6.3904 | 5.0774 | 3.7681 | 6.3867 |
| 354 | 4.7286 | 3.4054 | 6.0519 | 4.7334 | 3.4241 | 6.0428 |
| 355 | 4.5431 | 3.2199 | 5.8663 | 4.5432 | 3.2339 | 5.8525 |
| 356 | 5.1298 | 3.8065 | 6.4530 | 5.1340 | 3.8247 | 6.4434 |
| 357 | 5.4154 | 4.0922 | 6.7387 | 5.3290 | 4.0197 | 6.6384 |
| 358 | 5.5640 | 4.2408 | 6.8872 | 5.5598 | 4.2505 | 6.8692 |
| 359 | 5.4724 | 4.1492 | 6.7957 | 5.4668 | 4.1574 | 6.7761 |
| 360 | 4.6711 | 3.3478 | 5.9943 | 4.6728 | 3.3635 | 5.9822 |

Obs=Observation
F. C.D= Forecast for complete data
F. M.D= Forecast for missing data

## 7. Theil's statistics for accuracy of the forecast

The accuracy of the forecast was examined by using the Theil's U test which compares the accuracy of ARIMA model to that of a naive model. It simply uses the actual value for the last time period $Y_t$ as a forecast for $\hat{Y}_{t+1}$ , the formula for Theils U is (4):

$$U = \frac{RMSE(ARIMA)}{RMSE(naive)}. \tag{9}$$

where (RMSE) is the Root of the Mean Squared Error as being mathematically defined in the Eq.(10).

$$RMSE = \sqrt{\frac{1}{n} \sum_{t=1}^{n} (e_t)^2}. \tag{10}$$

where n is the number of observations in the series and e is an error term The result as in the tables (6) and (7) for both models ARIMA $(1, 0, 0)(0, 1, 1)_s$ and Naive for the MSE and RMSE. Therefore, Theil's is equal $U = 0.720864$ for the complete data and equal $U = 0.726352$ for the data with missing observation. These were less than 1, which means the model chosen was a good model. Since, a Theil's U greater than 1.0 indicates that the forecast model is worse than the naive model; a value less than 1.0 indicates that it is better. The closer U to 0 the better model that we have [6]. We observed that the values approximately close to each other which means that the method used to estimate the missing data was convenient at least on the data which used in this paper.

Table 6

*Values of MSE and RMSE for the ARIMA and Nave for the complete data*

| Model(complete data) | MSE | RMSE |
|---|---|---|
| ARIMA$(1,0,0)(0,1,1)_s$ | 0.441654 | 0.66457 |
| Naive | 0.849915 | 0.921908 |

Table 7

*Values of MSE and RMSE for the ARIMA and Nave with missing observation*

| Model(missing data) | MSE | RMSE |
|---|---|---|
| ARIMA$(1,0,0)(0,1,1)_s$ | 0.429536 | 0.65539 |
| Naive | 0.814152 | 0.902304 |

## 8. Conclusion

This paper investigates the application of Box and Jenkins technique to predict monthly average for rainfall in Pinang station by using the suggested new method to estimate the missing value. Model parameters are estimated using Autoregressive Integrated Moving Average (ARIMA) model in a period from 1st Jan 1969 to 31st Dec 1997. The model was tested in forecasting with the observed monthly average data in the same period. It has been found the best estimated ARIMA model for forecasting monthly average rainfall is the ARIMA $(1,0,0)(0,1,1)_s$ model. We compared the result of this model in both datasets with and without missing data. The equations for the model without missing data is

$$z_t = 0.159z_{t-1} - a_t + 0.857a_{t-12}$$

and the model with missing data is

$$z_t = 0.178z_{t-1} - a_t + 0.854a_{t-12}$$

The result checked with respect to the Naive test, which the Theil's is equal $U = 0.72086$ for the first one and for the second one is equal $U = 0.726352$ that means the result is closed to each other, that is, ARIMA $(1,0,0)(0,1,1)_s$ was a good model . The results indicate that time series techniques can be used to develop highly accurate short term forecasts of the monthly average rainfall depend on the past observation for Pinang station.

## References

[1] GENCAY R., SELCUK F. and WHITCHER B. (2002). *An Introduction to Wavelets and other filtering methods in finance and economics*, Permissions Department, Harcourt, Inc.

[2] JAMES W. T. and KURTZ, T. G. (2007). *A Comparison of Univariate Time Series Methods for Forecasting Intraday Arrivals at a Call Center.* Said Business School, University of Oxford.

[3] JOHN C. B. and DAVID A. D. (2003). SAS for Forecasting Time Series, Second Edition. Cary, NC: Institute Inc.

[4] PANKRATZ A. (1983). *Forecasting with Univariate Box-Jenkins Models*, Wiley New York.

[5] PAOLO B., ALBERTO M. and ROBERTO R. (1996). *Forecasting of storm rainfall by combined use of radar, rain gages and linear models* **1**. Atmospheric Research,Vol. 42, issue 1-4, pp. 199-216

[6] PATRICIA E G. (1994). Introduction to Time Series Modeling and Forecasting in Business and Economics, Cgraw-Hill M. Inc.

[7] RICHARD T. B.,SFETSOS A. and SANG-KUCK CH.(2002). *Modeling and forecasting from trend-stationary long memory models with applications to climatology* **1**. International Journal of Forecasting, Vol. 18, issue 2 pp. 215-226.

[8] SABRY M.,ABD - EL - 1 1 LATIF H. ,YOUSIF S. and BADRA N.(2007). *Use of Univariate Box and Jenkins Time Series Technique in Rainfall Forecasting* **1**. Australian Journal of Basic and Applied Sciences, (4) pp. 386-394.

[9] SFETSOS A. and COONICK A. H.(1966). *Univariate and multivariate forecasting of hourly solar radiation with artificial intelligence techniques* **1**. A Solar Energy Vol. 68, No. 2, pp. 169-178.